\documentclass[11pt]{article}
\usepackage{amsmath,amssymb}
\usepackage{graphicx,color}

\def\({\left(}
\def\){\right)}

\newcommand{\de}{\partial}
\newcommand{\be}{\begin{equation}}
\newcommand{\ba}{\begin{eqnarray}}
\newcommand{\ea}{\end{eqnarray}}
\newcommand{\ee}{\end{equation}}

\newcommand{\f}{\frac}
\newcommand{\s}{\sqrt}

\newcommand{\ap}{\alpha}

\newcommand{\ddd}{\cdot\cdot\cdot}
\newcommand{\no}{\nonumber \\}
\newcommand{\ep}{\epsilon}

 \def\de{\partial}

 \def\f {\frac}
 
 \def\ap{\alpha}

 \def\ddd{\cdot\cdot\cdot}
 \def\no{\nonumber \\}

 \def\ep{\epsilon}

\newcommand{\bea}{\begin{eqnarray}}
\newcommand{\eea}{\end{eqnarray}}
\newcommand{\bA}{\begin{array}}
\newcommand{\eA}{\end{array}}
\newcommand{\bc}{\begin{center}}
\newcommand{\ec}{\end{center}}
\newcommand{\al}{\alpha}

\newcommand{\ra}{\rightarrow}
\newcommand{\del}{\partial}

\newcommand{\ie}{{\it i.e.}}
\newcommand{\eg}{{\it e.g.}}

\newcommand{\Nf}{${\cal N}{=}4$}

\newcommand{\No}{${\cal N}{=}1$}

\begin{document}

\begin{titlepage}
\thispagestyle{empty}

\begin{flushright}
YITP-12-103
\\
IPMU12-0230
\\
\end{flushright}

\begin{center}
\noindent{\Large \textbf{$AdS$ plane waves and entanglement entropy}}\\
\vspace{2cm}

K. Narayan,
Tadashi Takayanagi
and Sandip P. Trivedi
\vspace{1cm}

{\it
 $^{a}$Chennai Mathematical Institute,\\
SIPCOT IT Park, Siruseri 603103, India.\\
\vspace{0.2cm}
 $^{b}$Yukawa Institute for Theoretical Physics,
Kyoto University, \\
Kitashirakawa Oiwakecho, Sakyo-ku, Kyoto 606-8502, Japan.\\
\vspace{0.2cm}
 $^{c}$Department of Theoretical Physics, Tata Institute of Fundamental
Research,\\ Homi Bhabha Road, Colaba, Mumbai 400005, India.
 }

\vskip 2em
\end{center}

\begin{abstract}
$AdS$ plane waves describe simple backgrounds which are dual to
anisotropically excited systems with energy fluxes.  Upon dimensional reduction, they reduce to hyperscaling violating spacetimes: in particular, the $AdS_5$ plane wave is known to exhibit logarithmic behavior of the entanglement entropy.
In this paper, we carry out an extensive study of the holographic entanglement entropy for strip-shaped subsystems in AdS plane wave backgrounds. We find that the results depend crucially on whether the strip is parallel or orthogonal to the energy current. In the latter case,
we show that there is a phenomenon analogous to a phase transition.
\end{abstract}

\end{titlepage}

\newpage

\section{Introduction}

The entanglement entropy is a very useful quantity which characterizes the ground state
of a quantum many-body system. It can be used as  a  quantum order parameter to
 distinguish
various quantum phases. Recently, entanglement entropy
has been actively explored both from the string theory side
and the condensed matter theory side. For example,
the properties of entanglement entropy of ground states
have been studied from the field theoretic
approach \cite{CCreview}
and the holographic approach \cite{HEEreview}.
However, our understanding of quantum entanglement
of excited states in quantum many-body systems is still limited at present.

It is well-known that the entanglement entropy correctly measures
the amount of quantum entanglement only for pure states. On the
other hand, if we compute the entanglement entropy for a finite
temperature system as an example of mixed state, it includes a
contribution from the thermal entropy in addition to that from the
quantum entanglement. Therefore we need to look at pure excited
states in order to estimate the quantum entanglement by employing
the  entanglement entropy. One such example is an excited state
created by a quantum quench, which is typically induced by a sudden
shift of parameters in a given quantum system. The entanglement
entropy increases under time evolution after the quantum quench.
In two dimensional conformal field theories (CFTs), some analytical
results have been known \cite{Quench}. Holographic calculations have
been done for any dimension \cite{HRT,Quenchhol}. In these
examples, the systems are excited homogeneously and isotropically.
Since the holographic description of quantum quenches requires complicated
numerical calculations of black hole formation in general, it is
quite difficult to get analytical results on the behavior of
entanglement entropy\footnote
{Nevertheless, a universal relation which is analogous to thermodynamics has been found recently in \cite{BNTU} for a small size subsystem.}.

The purpose of this paper is to study entanglement entropy in a more
tractable setup. Specifically we consider a CFT with a constant energy
flux $T_{++}$. This offers us a simple model of anisotropically
excited states with energy flow. It is holographically dual to an
AdS space with a plane wave \cite{Narayan:2012hk,Singh:2012un,Narayan:2012wn}\
(similar solutions in the form of shock waves in $AdS$ have been
studied previously \eg\ \cite{Janik:2005zt,Grumiller:2008va,Horowitz:1999gf}).
To define the entanglement entropy we need to specify a
subsystem which we trace out. We choose the subsystem to be a strip
with a finite width. Interestingly we will find that the behavior of
the holographic entanglement entropy (HEE) crucially depends on the
direction of the strip.

After a light-like compactification, as
shown in \cite{Narayan:2012hk,Singh:2012un,Narayan:2012wn}, these AdS
plane wave backgrounds become gravity duals with hyperscaling violation:
the $AdS_5$ plane wave in particular gives a spacetime exhibiting
logarithmic behavior of entanglement entropy.
Such a background can be realized in effective Einstein-Maxwell-Dilaton
theories \cite{Goldstein:2009cv,Cadoni:2009xm,
Charmousis:2010zz,Perlmutter:2010qu,Gouteraux:2011ce,Bertoldi:2010ca,
Iizuka:2011hg,Alishahiha:2012cm,Bhattacharya:2012zu,Kundu:2012jn} and has been
expected to be dual to non-Fermi liquids \cite{OTU,Sa}\ (see also
\cite{Dong:2012se,Narayan:2012hk,Kim:2012nb,Singh:2012un,
Dey:2012tg,Perlmutter:2012he,Ammon:2012je,Kulaxizi:2012gy,Narayan:2012wn}
for string realizations and further discussion).

The paper is organized as follows. In section 2, we review $AdS$ plane
wave backgrounds.  In section 3, we present results of holographic
calculations of entanglement entropy in the AdS plane wave for the case
where the strip subsystem is parallel with the energy current. In
section 3, we study the holographic entanglement entropy when the
subsystem is orthogonal to the energy current and show that there is a
behavior analogous to a phase transition. In section 5, we summarize
our conclusions.

\section{Reviewing AdS Plane Waves}

The gravity/5-form sector of IIB string theory contains the $AdS_5$ plane
wave \cite{Narayan:2012hk} (see also \cite{Singh:2012un,Narayan:2012wn})
\be\label{AdSplanewaveHom}
ds^2 = {R^2\over r^2} [-2dx^+dx^- + \sum_{i=1}^2 dx_i^2 + dr^2] + R^2 Q r^2 (dx^+)^2
+ R^2 d\Omega_5^2\ ,\qquad R^4\sim g_{YM}^2N{\al'}^2\ ,
\ee
as a solution with no other sources, with $Q$ a parameter of dimension
(boundary) energy density, and $d\Omega_5^2$ being the metric on $S^5$
(or other Einstein space).
Equivalently, the 5-dim part of the metric is a solution to
$R_{MN}=-{4\over R^2} g_{MN}$ arising in the effective 5-dim gravity
system with negative cosmological constant: the $g_{++}$ deformation
is a normalizable one. From the holographic stress tensor calculation,
one obtains
\be
T_{++}= {Q\over 4\pi G_5}\ .
\ee
Thus we see that this spacetime is dual to an excited state in the
boundary \Nf\ SYM conformal field theory with uniform constant
lightcone momentum density turned on. These spacetimes correspond
to a (chiral) wave on the boundary. Imposing a null energy
condition, we have $T_{++}\sim Q\geq 0$.

This spacetime (\ref{AdSplanewaveHom}) can be obtained
\cite{Singh:2012un} as a ``zero temperature''
double-scaling limit of boosted black D3-branes, using \eg\
\cite{Maldacena:2008wh}: to elaborate, consider black D3-branes
(\ie\ $AdS_5$ Schwarzschild spacetimes) with metric\
$ds^2=r^2 [-(1-r_0^4r^4) dt^2 + dx_3^2 +\sum_{i=1}^2 dx_i^2]
+ {dr^2\over r^2 (1-r_0^4r^4)}$ \
%(with boundary at $w=\infty$ and horizon at $w=r_0$)
and define $t={x^++x^-\over\sqrt{2}},\ x_3={x^+-x^-\over\sqrt{2}}$,
with lightcone coordinates $x^\pm$:\ after boosting by $\lambda$ as\
$x^\pm\ra \lambda^{\pm 1}x^\pm$, we obtain
%(and redefining to $r={1\over w}$ with boundary at $r=0$)
\be\label{blackD3lightcone0}
ds^2 = {1\over r^2} \left[-2dx^+dx^-+ {r_0^4 r^4\over 2}
(\lambda dx^++\lambda^{-1} dx^-)^2 +\sum_{i=1}^2 dx_i^2\right]
+ {dr^2\over r^2 (1-r_0^4r^4)}\ .
\ee
Note that  we have set the AdS radius $R$ to unity here. 
Now in the double scaling limit $r_0\ra 0,\ \lambda\ra\infty$, with
$Q={r_0^4\lambda^2\over 2}$ fixed, (\ref{blackD3lightcone0}) reduces
to (\ref{AdSplanewaveHom}).\ (For $r_0=0$, this is just $AdS_5$ in
lightcone coordinates.)\ Rewriting this in terms of just $Q, r_0$,
%with $\lambda^2={2Q\over r_0^4}$
gives
\be
\label{blackD3lightcone}
ds^2 = {1\over r^2} \left[-2dx^+dx^-+ Q r^4
\big(dx^+ + {r_0^4\over 2Q} dx^-\big)^2 +\sum_{i=1}^2 dx_i^2\right]
+ {dr^2\over r^2 (1-r_0^4r^4)}\ .
\ee
From \cite{Maldacena:2008wh}, we see that we now have other
energy-momentum components also turned on,
\be
T_{++}\sim \lambda^2 r_0^4\sim Q\ ,\qquad
T_{--}\sim {r_0^4\over\lambda^2}\sim {r_0^8\over Q}\ ,
\qquad T_{+-}\sim r_0^4\ ,\qquad T_{ij}\sim r_0^4\delta_{ij}\ .
\ee
Turning on a small $r_0$ about (\ref{AdSplanewaveHom}), this means
$T_{++}$ is dominant while the other components are small. In some
sense, this is like a large left-moving chiral wave with $T_{++}\sim Q$,
with a small amount of right-moving stuff turned on. This spacetime
interpolates between the usual unboosted black D3-brane ($\lambda=1$, \ie\
$Q={r_0^4\over 2}$) and the $AdS_5$ plane wave ($\lambda\ra\infty$,
with $Q$ fixed). There are two nontrivial scales here, $Q$ and $r_0$:
for small $r_0$, we expect that physical observables such as
entanglement entropy are dominated by the $AdS$ plane wave limit,
\ie\ by $Q$, with small $r_0$-dependent corrections.

Let us now consider $x^+$-dimensional reduction of
(\ref{AdSplanewaveHom}) as in \cite{Narayan:2012hk}: the 5-dim part
of this metric can be rewritten (relabelling $x^-\equiv t$) as,\
$ds^2 = R^2( -{dt^2\over Q r^6} + {\sum_{i=1}^2 dx_i^2 + dr^2\over r^2}
+ Q r^2 (dx^+ - {dt\over Q r^4})^2)$ .\
This gives the effective (bulk) $3+1$-dim Einstein metric and exponents
\be\label{liftheta=d-1}
ds^2_E\ %=\ (R^2Qr^2)^{1/(4-2)}
%R^2 \left( -{dt^2\over Qr^6} + {dx_i^2 + dr^2\over r^2} \right)
=\ {R^3\sqrt{Q}\over r} \Big(-{dt^2\over Qr^4} + \sum_{i=1}^2 dx_i^2 + dr^2\Big)\ ,\qquad
 \ \theta=1 ,\ \ z=3\ ,
\ee
along with an electric gauge field\ $A=-{dt\over Qr^4}$ and scalar\
$e^\phi\sim r$. Above, we have compared with the hyperscaling
violating metric in the form\
$ds^2= r^{2\theta/d_i} (-{dt^2\over r^{2z}} + {\sum_{i=1}^2 dx_i^2+dr^2\over r^2})$,
with boundary spatial dimension $d_i$.
%We have retained the nontrivial scales $R, Q$ to illustrate their
%higher dimensional origin.
%This dimensionally reduced metric has ``boundary'' spatial dimension $d=2$ and is of the form (\ref{lifhyper}) with
%\be
%2(1-{\theta\over 2}) = 1\ \Rightarrow\ \theta = 1 = d-1\ ,\qquad
%2(z-1)=4 \Rightarrow\ z=3\ .
%\ee
This lies in the hyperscaling violating family ``$\theta=d_i-1$''
exhibiting logarithmic behavior of entanglement entropy (as we will review later)
and has thus been argued to correspond to a gravitational dual
description of a theory with hidden Fermi surfaces
\cite{OTU,Sa}% Ogawa:2011bz,Huijse:2011ef}
(see also \cite{Dong:2012se}).
It is thus interesting to explore these $AdS$ plane waves further,
in particular from the higher dimensional point of view.

More generally, we have the (purely gravitational)
$AdS_{d+1}$ deformation, which is the $AdS_{d+1}$ plane wave,
\be\label{met}%\label{AdSd+1nullNorm}
ds^2 = {R^2\over r^2} \left(-2dx^+dx^- +\sum_{i=1}^{d-2} dx_i^2 + dr^2\right)
+ R^2Qr^{d-2} (dx^+)^2\ ,
\ee
the $x_i$ being $(d-1)$-dim (boundary) spatial coordinates,
with $Q$ a parameter of energy density in $d$-dimensions.
This is a solution to\ $R_{MN}=-{d\over R^2} g_{MN}$, \ie\ to gravity
with a negative cosmological constant.
It is also useful to introduce the time and space coordinate
$(t,x_{d-1})$ such that
\be
x^\pm=\f{t\pm x_{d-1}}{\s{2}}\ .
\ee
This can then be used to obtain the spacetime (\ref{met}) %{AdSd+1nullNorm})
as a double-scaling limit of the $AdS_{d+1}$ black brane in
%Similar interpretations exist for $AdS_{d+1}$ black branes in
lightcone coordinates near a double-scaling limit \cite{Singh:2012un},
resulting in a near-extremal $AdS_{d+1}$ plane wave: most of our
discussion with the regulated (finite temperature) case will be for
the $AdS_5$ plane wave.

%This metric exhibits the scaling symmetry
%$x_i\ra \lambda x_i ,\quad r\ra \lambda r ,\quad x^-\ra \lambda^{2+d/2} x^- ,
%\quad x^+\ra \lambda^{-d/2} x^+$ .
Now, dimensionally reducing on the $x^+$-dimension (and relabelling
$x^-\equiv t$) gives the metric and exponents
\be\label{lifHypd}
ds_E^2 %= R^2 (R^2Qr^{D-3})^{1/(D-3)}
%\left( -{dt^2\over Qr^{D+1}} + {dx_i^2 + dr^2\over r^2} \right)
=\ {R^2 (R^2Q)^{1/(d-2)}\over r}
\Big(-{dt^2\over Qr^{d}} +\sum_{i=1}^{d-2} dx_i^2 + dr^2\Big)\ ,\qquad
z={d-2\over 2}+2\ ,\quad \theta={d-2\over 2}\ .
\ee
%The dimensionally reduced metric above has ``boundary'' spatial
%dimension $d-1$ and is of the form (\ref{lifhyper}) with
%\be\label{AdsDztheta}
%z={d\over 2}+2\ ,\qquad \theta={d\over 2}\ .
%\ee
For the special case of $d-2=2$, this $\theta$ value lies in the
special family ``$\theta=d_i-1$''\ as we have seen above\footnote{Note
that the $d_i$ in the expression $\theta=d_i-1$ is the boundary spatial
dimension, while we are discussing $AdS_{d+1}$ plane waves with $d$
the boundary spacetime dimension in the higher dimensional description.}.
From the lower dimensional point of view (the ``$\#$'' are numerical
constants), the $d$-dim action\
$\int d^{d+1}x \sqrt{-g^{(d+1)}}\ (R^{(d+1)} - 2\Lambda)$ dimensionally
reduces as
\be
\int dx^+ d^{d}x \sqrt{-g^{(d)}}\ (R^{(d)} - \#\Lambda e^{-2\phi/(d-2)}
- \# (\del\phi)^2 - \# e^{2(d-1)\phi/(d-2)} F_{\mu\nu}^2 ) ,\qquad
\ee
where the scalar is\ $g_{d+1,d+1} = e^{2\phi}$, the (purely electric)
gauge field is\ $A=-{dt\over r^{d}}$\ and the $d$-dimensional
metric undergoes a Weyl transformation as\
$g^{(d)}_{\mu\nu}=e^{2\phi/(d-2)} g^{(d+1)}_{\mu\nu}$. It is
straightforward to check that the solution (\ref{lifHypd}) is
consistent with the equations of motion, with the scalar of the
form\ $e^{2\phi}=r^{d-2}$.

There is a more general family of $AdS$ null deformations
\cite{Narayan:2012wn} which include inhomogenous $AdS$ plane waves, of
the form\ %\be\label{AdSnullgen}
$ds^2 = {1\over r^2} [-2dx^+dx^- + \sum_{i=1}^{2} dx_i^2 + dr^2] + g_{++} (dx^+)^2 +
d\Omega_5^2$ ,\ with $g_{++}(r,x_i)$ spatially varying. This also
includes the case $g_{++}\xrightarrow{r\ra 0} const=K$ , sourced by
other matter fields, which are asymptotic to $z=2$ 4-dim Lifshitz
spacetimes upon $x^+$-dimensional reduction \cite{z=2Lif}. All these
solutions, including (\ref{AdSplanewaveHom}), have finite curvature
invariants everywhere, which are the same as those for $AdS_5\times
S^5$. Furthermore, they all preserve some supersymmetry. Due to the
lightlike nature, nonzero contractions involving curvature components
vanish (this can be checked explicitly for low orders) and thus these
spacetimes are likely to be $\al'$-exact, somewhat analogous to plane
waves and $AdS_5\times S^5$. This makes them possibly more interesting
as string backgrounds.

It is worth noting the possibility of tidal forces diverging in the
deep interior $r\ra\infty$, even though curvature invariants are
regular and the same as in $AdS_5\times S^5$: this is a general
feature of plane wave (and in fact any lightlike) spacetimes.  This
may reflect the fact that from the lower dimensional point of view, we
have a hyperscaling violating spacetime, which is conformal to
Lifshitz, and there are curvature singularities arising from the
conformal factor (unlike Lifshitz spacetimes which only have diverging
tidal forces); see \eg\ \cite{Copsey:2012gw} for a recent
discussion. From the field theory point of view, this is the question
of whether turning on uniform $T_{++}$ density has certain
pathologies. From the bulk point of view, this sort of a singularity
if it exists is often regarded as mild, possibly regulated by finite
temperature effects. In this sense, thinking of the homogenous $AdS$
plane wave as a zero temperature ``chiral'' limit of the boosted black
brane is useful: physically any singularity will be cloaked by the finite
temperature horizon. From the dual point of view, we are considering
a certain boosted limit of thermal states in the CFT: for finite (if
large) boost, we expect this is well-behaved.

\subsection{Entanglement Entropy for Light-like Subsystems in AdS$_5$ Plane Wave}

The entanglement entropy $S_A$ for the subsystem $A$ is defined by
$S_A=-\mbox{tr}\rho_A\log\rho_A$, where $\rho_A$ is defined by tracing out
the density matrix $\rho$ for the total system over the subsystem $B$, which is the complement
of $A$, i.e. $\rho_A=\mbox{tr}_B\rho$. In any backgrounds of AdS/CFT,
we can holographically calculate the entanglement
entropy $S_A$ from the area of extremal surface $\gamma_A$  which ends on the boundary of $A$. An extremal surface is defined by the one whose area functional gets stationary under any infinitesimal deformations with a fixed boundary condition. The area is calculated in terms of
the Einstein frame. If there are several extremal surfaces, we need to pick up the one with the minimal area among them. This is called the (covariant) holographic entanglement entropy (HEE) \cite{HRT}:
\be
S_A=\mbox{Min}_{\gamma_A\in ES.}\left[\f{\mbox{Area}(\gamma_A)}{4G^{(d+1)}_N}\right], \label{covee}
\ee
where $G^{(d+1)}_N$ is the Newton constant in the $d+1$ dimensional gravity theory we consider.
In static backgrounds, since we can restrict to a time slice, $\gamma_A$ is reduced to a minimal area surface on that slice and leads to the minimal surface prescription \cite{RT}.
In many parts of this paper we will omit the factor $\f{1}{4G^{(d+1)}_N}$ because we are
mainly interested in the dependence on the size of the subsystem and thus the overall factor is not important for our purpose.

We now review the logarithmic violation of the area law for entanglement entropy in the $AdS_5$ plane wave, using the HEE (\ref{covee}) and finding the area of a bulk
minimal surface bounding a subsystem $A$ in the shape of a strip in
the $x_1,x_2$-plane.
From the point of view of the lower dimensional theory obtained by
$x^+$-dimensional reduction, it is natural to consider subsystems
$A$ that extend along the $x^+$-direction completely, and the
corresponding bulk minimal surface lying on a constant lightcone
time $x^-$ slice (which corresponds to a constant time slice since
$x^-\equiv t$).
%The spatial metric on such a slice is\
%$ds^2 = {R^2\over r^2} (dx_i^2 + dr^2) + R^2Qr^2 (dx^+)^2$.\
%This then gives in the bulk a minimal surface wrapping the
%$x^+$-direction and bounding the subsystem $A$ in the $x_i$-plane.
%, as is clear from the spatial metric above.

Consider a strip region in the $x_1$-direction given by
$-l\leq x_1\leq l$, extending along the $x_2$-direction: the minimal
surface is parametrized by $x=x(r)$, and its area gives the
entanglement entropy
\be\label{HolEE}
S_A = {1\over 2G^{(5)}_N} \int_0^L {Rdy\over r} \int_0^{L_+} \sqrt{g_{++}} dx^+
\int {R\sqrt{dx^2 + dr^2}\over r} \
=\ {L L_+R^3\sqrt{Q}\over 2G^{(5)}_N} \int_\epsilon^{r_*} dr\
{\sqrt{1 + (x')^2}\over r} \ ,
\ee
where $\epsilon$ is the near-boundary cutoff (\ie\ the UV cutoff in
the field theory). The minimal surface has a turning point $r_*$
where ${dr\over dx}|_{r_*}=0$.
The minimal surface then is given by the half circle:
\be
x=\sqrt{l^2-r^2},
\ee
and thus we can estimate as
\be\label{logviol}
S_A =\ {L_+R^3\sqrt{Q}\over 2G^{(5)}_N}\ L \log {l\over\epsilon}\ .
\ee
Using $G^{(4)}_N={G^{(5)}_N\over L_+}$, this gives the logarithmic
behavior, as expected from the lower dimensional theory. We have
effectively taken $l\gg \epsilon$, so that the strip width $l$ is
macroscopic relative to the UV cutoff $\epsilon$ in the field theory.
When the strip size shrinks to roughly the cutoff, we have a
cross-over to the UV behaviour in the field theory: in this case, we
expect the entanglement entropy for $AdS_5$ in lightcone time slicing
which vanishes, as vindicated by (\ref{logviol}) for $l\sim\epsilon$.
As $Q\ra 0$, this surface degenerates and becomes null, and the
corresponding area vanishes.

Note that this calculation (\ref{logviol}) above arises from just
the 5-dim part of the spacetime, so it also applies to $AdS_5\times X^5$
plane waves dual to various \No\ super Yang-Mills theories.

\section{Entanglement Entropy in AdS Plane Wave: Case A}

The purpose of this paper is to give an extensive study of holographic entanglement entropy in the AdS plane wave background (\ref{met}) and its regularized
one. Especially we will look at the entanglement
entropy $S_A$ when the subsystem $A$ is given by a (infinitely
extended) strip with a finite width $l$ at a constant time.
Then the entanglement entropy
consists of two terms: one is the area law divergence \cite{Area} and the other
is the finite term \cite{CH,RT}. We expect that the area
law divergence remains the same as that for the pure AdS dual to the ground state,
while the subleading finite term will be modified in our excited backgrounds.

\begin{figure}[t]
   \begin{center}
     \includegraphics[height=6cm]{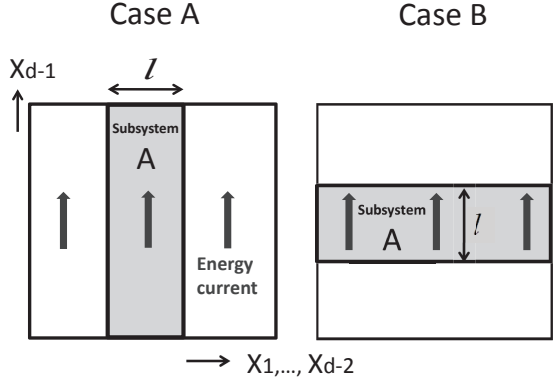}
      \end{center}
   \caption{The two different choices of the strip subsystem $A$. The gray arrows represents the energy current. The left and right are called case A and case B, respectively.}\label{fig:sub}
\end{figure}

There are two different choices of such systems depending on whether the finite width direction is (a) either of $(x_1,x_2,\ddd, x_{d-2})$, called case A or (b) $x_{d-1}$, called case B (see Fig.\ref{fig:sub}). We will study the case A below and the case B in the next section because
in the latter case, a more careful analysis is needed. Notice that in case A, the energy current is parallel with the strip, while it is orthogonal in case B.

\subsection{Holographic Analysis in Case A}

In case A, the width direction is orthogonal to
$x_+$ and $x_-$. We can take it to be along $x_1$. Thus we need to assume $d\geq
3$. Then the subsystem $A$ is specified by \be 0\leq x_1\leq l,\ \ \
(x_+,x_-)=(\ap y,-\beta y), \ \ \ -\infty<y,x_2,x_3,\ddd,x_{d-2}
<\infty. \ee
The extremal surface $\gamma_A$ is specified by the function
$x_1=x(r)$. We define the maximal value of $r$ to be $r_*$, where $\gamma_A$ turns around, changing the sign of $\f{dr}{dx}$.

The area functional is given by \be {\mbox{Area}}=R^{d-1}V_{d-2}\cdot
2\s{\ap}\int^{r_*}_\ep dr r^{-(d-1)}\s{(1+x'^2)(2\beta+Qr^{d}\ap)},
\ee where $V_{d-2}$ denotes the (infinite) volume
in the $y$ and $(x_2,\ddd,x_{d-2})$ direction. The infinitesimally small parameter
$\ep$ represents the UV cut off.

The extremal surface is specified by the function $x_1=x(r)$.
 The equation of motion leads to \be \de_r x=\f{Ar^{d-1}}{\s{2\beta
 +Q\ap r^d-A^2r^{2(d-1)}}}, \label{cam}
\ee where $A$ is the integration constant. Note that at $r=r_*$, the denominator of the right hand side of (\ref{cam}) vanishes. Thus we find \be
l=2\int^{r_*}_0 dr \f{Ar^{d-1}}{\s{2\beta
 +Q\ap r^d-A^2r^{2(d-1)}}}. \ee

Now the area functional is rewritten as \be {\mbox{Area}}=
2\s{\ap}V_{d-2}R^{d-1}\int^{r_*}_\ep \f{dr}{r^{d-1}}\cdot \f{2\beta+Q\ap
r^{d}}{\s{2\beta+Q\ap r^d-A^2 r^{2(d-1)}}}. \label{cona} \ee

\subsection{Behavior of Entanglement Entropy}

The entanglement entropy $S_A$ in a quantum field theory
is defined for a subsystem $A$ which is on a constant time slice.
This corresponds to the choice $\beta=\ap=1$.
In this case, the leading divergence is the standard area law \be S_A\sim
O(\ep^{-(d-2)}). \ee The effects due to non-vanishing flux $Q$  turn
out to be finite. When $Q$ and $l$ are both large, we find $l\sim r_*$ and $Qr_*^{d}\simeq A^2
r^{2(d-1)}_*>>1$.  In this case, we find that the finite part of
$S_A$ behaves like\footnote{Here we omitted factors like $\f{1}{G^{(d+1)}_N}$ as we are not interested in the overall factor which is fixed once the CFT is given.}
\be
S_A|_{finite}\sim \pm V_{d-2}\s{Q} \cdot l^{2-\f{d}{2}}. \label{eesl} \ee
The sign in front of (\ref{eesl}) is $+$ for $d<4$ and $-$ for $d>4$. In the case $d=4$ we need to
replace (\ref{eesl}) with
\be
S_A|_{finite}\sim V_{2}\s{Q}\cdot\log\left(l\cdot Q^{\f{1}{4}}\right). \label{loge}
\ee

In general, when $l$ is large, the above result is smaller than that of the thermal entropy, which is extensive and is proportional to $l$, while it is larger than the finite
contribution in CFTs at zero temperature, which scales like $\sim O(l^{-(d-2)})$. Moreover,
the finite contribution is a monotonically increasing function of $l$. This agrees with our
intuitive expectation: the active degrees of freedom should increase
because the energy flow excites the system.

We can also consider disconnected extremal surfaces defined by
$x(r)=$const., which extend from $r=\ep$ to the deep IR limit $r=\infty$.
However, they always give larger areas and therefore
do not contribute. When $d\leq 4$, it is clear that the IR contributions to
the areas diverge as $\int^\infty_\ep dr r^{1-d/2}=\infty$. When $d>4$, their contributions are finite, but numerically we could confirm that they are larger than the connected area (\ref{cona}). This situation is similar to the holographic calculations in the pure AdS spaces.

Two more  comments are worth making. First, when $d=3$, so that one is 
dealing with the $AdS_4$ plane wave and a $2+1$ dimensional boundary 
theory, the finite part of $S_A$ in eq.(\ref{eesl}) becomes
\begin{equation}
\label{threedim}
S_A|_{finite}\sim V_1 \sqrt{Q} \sqrt{l}
\end{equation}
where $V_1$ is the length of the region in the $y$ direction. 
We see that the log term in $d=4$ is replaced by  a power law enhancement going like $\sqrt{l}$. 
 While examples of string constructions or constructions in gauged supergravity 
giving rise to hyperscaling violating geometries are well known, to our knowledge, the  AdS plane wave
is the first example of a string construction which  gives rise to  such a power law enhancement.

Second, note that the AdS plane wave solution represents a valid state in 
any CFT which admits a smooth gravity dual since it arises simply from the 
AdS Schwarzschild solution in the limit of infinite boost. 
As  such, our result for the entanglement entropy in  eq.(\ref{eesl}) 
is also quite general applying, for example  
in the $d=4$ case to the \Nf\ SYM theory, the Klebanov-Witten theory 
from D3-branes on the conifold, and more general \No\ superconformal 
Yang-Mills theories dual to $AdS_5\times X^5$ (where $X^5$ is the 
Sasaki-Einstein base space of the Calabi-Yau cone where the D3-branes 
are stacked), and in the $d=3$ case to the M2 brane theory, ABJM theory etc.

\subsection{Light-like Limit}

The limit where $A$ gets light-like is given by $\beta=0$. This corresponds to a generalization
of entanglement entropy and its precise definition in quantum field theory can be given by the replica method at least formally (analogous to \cite{Quench}).

We can also set $\ap=1$
without losing generality. Then we find $l\sim r_*$. Thus it is clear that the leading term with respect to $\ep(\to 0)$ behaves \be
S_A\sim V_{d-2}\s{Q}\cdot O(\ep^{2-\f{d}{2}}). \label{eeelt} \ee Thus, $S_A$ is finite
for $d< 4$. For $d=4$, $S_A$ is logarithmically divergent \cite{Narayan:2012hk}
\be
S_A\sim V_{2}\s{Q}\cdot \log\f{l}{\ep}, \label{eelt}
\ee
as we reviewed in subsection 2.1. Note that the divergence is always
smaller than the space-like case $\beta\neq 0$.

\subsection{Analysis in Regularized AdS Plane Wave}

The above calculations can be repeated for the regulated or
near-extremal $AdS_5$ plane wave (\ref{blackD3lightcone}), with
finite boost $\lambda$. We obtain
\be
S_A\ \sim\ V_2 \int {dr\over r^3} \sqrt{(\del_rx)^2+{1\over 1-r_0^4r^4}}
\sqrt{2\al\beta + Qr^4 (\al-{\beta r_0^4\over 2Q})^2}
\ee
giving
\be
\del_r x = {1\over 1-r_0^4r^4}  {pr^3\over \sqrt{2\al\beta + Qr^4
(\al-{\beta r_0^4\over 2Q})^2 - p^2r^6}}\ ,
\ee
and thus
\be
S_A\ \sim\ V_2 \int {dr\over r^3} {2\al\beta + Qr^4
(\al-{\beta r_0^4\over 2Q})^2\over \sqrt{1-r_0^4r^4}
\sqrt{2\al\beta + Qr^4 (\al-{\beta r_0^4\over 2Q})^2 - p^2r^6}}\ .
\ee
For $\beta=0$, we have a lightlike surface, giving\
$S_A\ \sim\ V_2\sqrt{Q} \int {dr\over r}
{1\over\sqrt{(1-{p^2r^2\over Q\alpha^2})(1-r_0^4r^4)}}$\ with a logarithmic leading
divergence as before. In addition we have a finite subleading
piece $~ V_2 \sqrt{Q} {r_0^4\over l^4}$ which is cutoff independent.
For large size, we expect that the surface wraps part of the
horizon.

For a spacelike surface, we can take $\al=\beta=1$ as before, and
the leading divergence reflects the area law. Taking $r_0$ small, we
expect the extremal surface to be essentially governed by the $AdS$
plane wave background, with corresponding entanglement entropy.

As a check, note that in the limit where $Q={r_0^4\over 2}$ \ie\ the
familiar black D3-brane, taking $\al=\beta=1$, we recover the
familiar expression\
$S_A\ \sim\ V_2 \int {dr\over r^3} {2\over \sqrt{1-r_0^4r^4}
\sqrt{2 - p^2r^6}}$\ for the black D3-brane.

\section{Entanglement Entropy in AdS Plane Wave: Case B}

Next we study the case $B$ where the width direction of the strip $A$ is parallel to
$x_{d-1}$.
More generally we assume that the strip $A$ is defined by \be
-\f{\Delta x^+}{2}\leq x^+ \leq \f{\Delta x^+}{2},\ \ \ -\f{\Delta
x^-}{2}\leq x^- \leq \f{\Delta x^-}{2},\ \ \ -\infty<x_i<\infty. \ee
We will denote the regularized length in each of the $x_i$
directions by $L(>>l)$. The dimension $d$ is assumed to be
$d\geq 2$.

In this background, we again apply the covariant holographic entanglement
entropy (\ref{covee}) to our non-static spacetime (\ref{met}). The
extremal surface $\gamma_A$
can be specified by \be x^+=x^+(r),\ \ \ x^-=x^-(r). \ee

The area functional looks like \be {\mbox{Area}}=2R^{d-1}V_{d-2}\int^{r_*}_\ep
\f{dr}{r^{d-1}} \s{1-2(\de_r x^+)(\de_r x^-)+Qr^{d}(\de_r x^+)^2},
\ee
where $\ep$ is the UV cut off as before.

The equation of motion leads to \ba &&
\f{r^{d-1}}{A}=\f{Qr^d(\de_r x^+)-\de_r x^-}{\s{1-2(\de_r x^+)(\de_r
x^-)+Qr^{d}(\de_r x^+)^2}} ,\no  && \f{r^{d-1}}{AB}= \f{\de_r
x^+}{\s{1-2(\de_r x^+)(\de_r x^-)+Qr^{d}(\de_r x^+)^2}}, \ea where
$A$ and $B$ are integration constants and we assume $A>0$ and $B>0$.

This leads to \ba \de_r
x^+=\f{1}{\s{\f{A^2B^2}{r^{2(d-1)}}+Qr^{d}-2B}},\no \de_r
x^-=\f{Qr^d-B}{\s{\f{A^2B^2}{r^{2(d-1)}}+Qr^{d}-2B}}. \ea The
turning point $r=r_*$ of the extremal surface is determined by \be
\f{A^2B^2}{r_*^{2(d-1)}}+Qr_*^{d}-2B=0. \label{zeroc} \ee

In this way, we obtain the relations \ba && \f{\Delta
x^+}{2}=\int^{r_*}_0
\f{dr}{\s{\f{A^2B^2}{r^{2(d-1)}}+Qr^{d}-2B}},\no && \f{\Delta
x^-}{2}=\int^{r_*}_0
\f{(Qr^d-B)dr}{\s{\f{A^2B^2}{r^{2(d-1)}}+Qr^{d}-2B}}. \label{delx}
 \ea
In order to take the subsystem $A$ to be space-like, we can
assume $\Delta x^+>0$ and $\Delta x^-<0$.

Finally, the area of the extremal surface is computed as \be\label{SEadspw}
\mbox{Area}=2R^{d-1}V_{d-2}\int^{r_*}_{\ep}\f{dr}{r^{d-1}}\cdot
\f{AB}{\s{A^2B^2-2Br^{2(d-1)}+Qr^{3d-2}}}. \ee

\subsection{Exact Analysis in $d=2$}

In the $d=2$ case, we can analytically perform the previous
integrations. This corresponds to a particular limit of the result
discussed in \cite{HRT}.

We obtain \ba && \Delta x^+=\f{1}{\s{Q}}\log \f{1+A\s{Q}}{\s{1-Q
A^2}},\no && \Delta x^-=-AB. \ea This can be solved into \be
A=\f{\tanh(\s{Q}\Delta x_+)}{\s{Q}}.\ee The length is given by \ba
\mbox{Length}&=&R\log \f{4}{\ep^2}+R\log \left|\f{w_*}{2-\f{2w_*}{A^2B^2}}\right|
\no &=& R\log\f{2}{\ep^2}+R\log\left[\f{-\Delta^-\sinh(\s{Q}\Delta
x^+)}{\s{Q}}\right]. \ea

Finally, the holographic entanglement entropy is found to be
\be
S_A=\f{c}{6}\log\left[\f{-2\Delta^-\sinh(\s{Q}\Delta
x^+)}{\ep^2\s{Q}}\right]. \label{chth}
\ee
We can confirm that this agrees with the result in CFTs by
 taking the chiral limit of the analysis in \cite{HRT}.

It may be interesting to define the light-like limit (i.e. the limit the subsystem $A$
gets closer to light-like) by \be \Delta x^- \Delta x^+ =-\ep^2, \ \
\ \ \ \ \ \Delta x^+=\mbox{finite}>0.  \ee

In this limit, $S_A$ becomes finite: \be
\f{6S_A}{c}=\f{\mbox{Length}}{R}=\log\f{2\sinh(\s{Q}\Delta x^+)}{\s{Q}\Delta x^+}\simeq
\log 2 +\f{Q}{6}(\Delta x^+)^2 + O(Q^2). \ee

\subsection{The analysis in $d\geq 3$}

Let us move on to the higher dimensional case. We concentrate on the case where
the subsystem  $A$ is on a fixed time slice ($t=$fixed) and thus we require
\be
\Delta x^+ =-\Delta x^{-}=\f{l}{\s{2}},  \label{delt}
\ee
where $l$ is the width of the subsystem $A$.

Assume that the width $l$ is very large.
The length $l$ gets infinitely large only when the denominator $\f{A^2B^2}{r^{2(d-1)}}+Qr^{d}-2B$ develops a double zero at $r=r_*$.  This condition and (\ref{zeroc}) can be solved at this degenerate point as follows
\ba\label{doublezeroAdSpw}
&& B=\f{3d-2}{4(d-1)}Qr^{d}_*, \no
&& A^2=\f{8(d-1)d}{(3d-2)^2}\cdot\f{r^{d-2}_*}{Q}.
\ea
However, if we plug this in the integral we find that $\Delta x^-$ gets {\it positively} divergent.
This means that in the limit $\Delta x^+\to +\infty$, we always find $\Delta x^-\to \infty$ and therefore the subsystem $A$ gets time-like, which is not what we want. Thus this tells us that
 there exists an upper bound for $l$, which scales like $Q^{-1/d}$ for this connected
extremal surface. Note also that we need to take $B>1$ to satisfy (\ref{delt}).
We find that the width is vanishing at $B=1$.

\begin{figure}[t]
   \begin{center}
     \includegraphics[height=4cm]{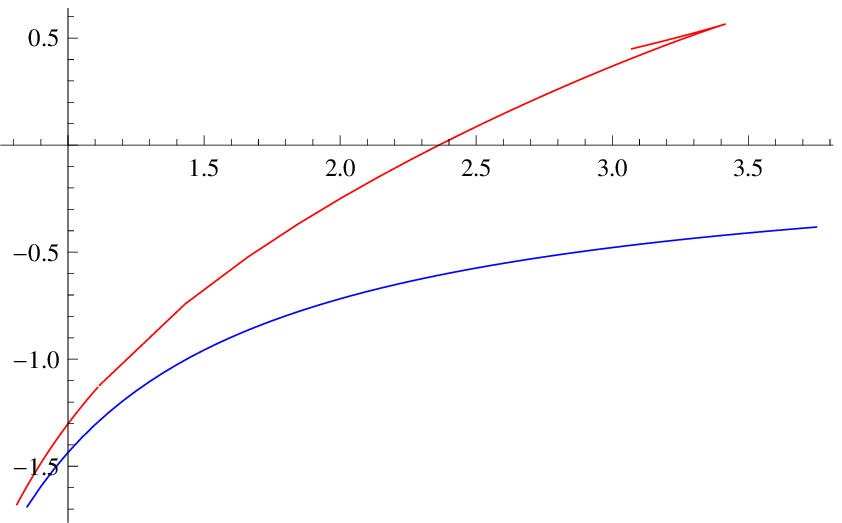}
   \hspace{1cm}
     \includegraphics[height=4cm]{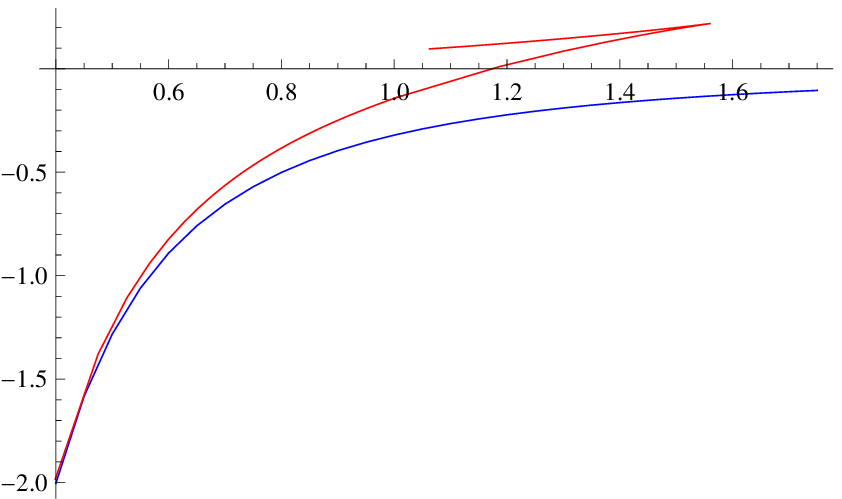}
      \end{center}
   \caption{Numerical plots of the regularized areas of extremal surfaces as functions of
    the width $l=\s{2}\Delta x^+=-\s{2}\Delta x^-$ of $A$. We set $Q=R=1$. The left and right graph corresponds to the results for $d=3$ and $d=4$, respectively. In each graph, the blue curve corresponds to the connected surface in the pure AdS space while the red one to the connected one in the AdS plane wave. The area is regularized by subtracting the area of the disconnected surface given by $x^\pm=$constant.}\label{fig:PHT}
\end{figure}

 Actually there is another candidate of the extremal surface. This is the disconnected surface simply given by $x^\pm=$const. We plotted areas as functions of the width $l/\s{2}\equiv \Delta x^+=-\Delta x^-$ in Fig.\ref{fig:PHT}, where we subtracted the disconnected surface area from
 the connected one. The covariant
holographic entanglement entropy is defined by choosing the one with the minimal area among
extremal surfaces as in (\ref{covee}). Thus our results in Fig.\ref{fig:PHT} show that there is a sort of phase transition at the width $l_c\sim 2.4$ for $d=3$ and $l_c\sim 1.2$ for $d=4$. For $l<l_c$ the connected surfaces are favored, while for
$l>l_c$ the disconnected ones are chosen. This behavior is a bit similar to the behavior in
the confining backgrounds such as AdS solitons \cite{NiTa,KKM}. Notice also that we can easily estimate $l_c\sim Q^{-1/d}$.

Naively, the existence of the energy flux
$T_{++}\propto Q>0$ just excites the system in a similar manner to finite temperature systems because the system is excited at the energy scale $Q^{1/d}$. Indeed, this speculation is true in the $d=2$ case as can be seen from (\ref{chth}). However in higher dimensions $d\geq 3$, as we found here, the result of holographic entanglement entropy is rather different from that in the finite temperature system, which has a positive and extensive (finite) contribution to $S_A$. Our results for $d\geq 3$, plotted in Fig.\ref{fig:PHT}, show that there are positive contributions, but they are not extensive at all. The presence of the phase transition is a special feature of case B, which does not appear in the case A.

It might be useful to note that in the $d=2$ case, the metric (\ref{met}) is equivalent to that of the extremal rotating BTZ black hole via a coordinate transformation. However, this is not true in higher dimensions. Therefore only $d=2$ has non-zero (thermal) entropy and it is very natural that we find the extensive behavior in this case. On the other hand, for $d\geq 3$, things are different and the systems are far from thermal states.

\subsection{An Interpretation of the different results between Case A and Case B}

It is clear that the difference between the case A and case B comes from the direction of energy current relative to the strip direction (see Fig.\ref{fig:sub}). When the energy flows along the strip (case A), the energy does not leak from the strip. Therefore, the system gets simply excited and the entanglement entropy increases monotonically with $l$
(see also \cite{BNTU} for the related argument in static systems). This intuitive expectation agrees with our result of HEE in section 3.

However, in case B, the strip extends in the
direction orthogonal to the energy flow and thus there is a constant energy exchange with adjacent regions. In this non-trivial setup, our holographic analysis shows that there is a maximal width $l_c\sim Q^{1/d}$ above which $S_A$ gets constant.

Since the system we consider
have the energy flux $T_{++}\sim Q$, each excited mode has the wave length of order $(T_{++})^{1/d}\sim Q^{1/d}$, which is analogous to the thermal screening length at finite temperature. Thus for a larger subsystem $l>l_c$, there is no correlation\footnote{
Indeed, in the replica method, the calculation of entanglement entropy can be regarded as the correlation function of two defects
which produce the deficit angles. In two dimensional CFTs, they are reduced to two point functions of twisted vertex operators \cite{Cardy}. We can also generalize this into higher dimensions in principle \cite{RT,HMSY}.} or entanglement between the deep inside of $A$ and $B$ except the regions near the boundary of $A$.
This explains why $S_A$ in the case B does not depend on $l$ when $l>l_c$. As usual, the large $N$ limit amplifies this phenomena and leads to the sharp phase transition.
On the other hand, if we consider the correlation in the direction orthogonal to the
 energy current as in case A, then there is no screening effect and thus there is no phase transition.

\subsection{Regularized $AdS$ plane wave: $x^\pm$-coordinates}

To understand better the phase transition above, let us regulate the
$AdS$ plane wave with the non-extremal background (\ref{blackD3lightcone})
containing a horizon at $r_0$: for concreteness, we consider regulating
the $AdS_5$ plane wave in terms of the boosted black D3-brane and
Case B, with the strip width parallel to $x_3$. Then the entanglement
entropy is given by
\be
S_A\ \sim\ V_2\int {dr\over r^3} \sqrt{{1\over 1-r_0^4r^4}
- 2(\del_rx^+)(\del_rx^-) + Qr^4 \big(\del_rx^+ +
{r_0^4\over 2Q} \del_rx^-\big)^2}\ .
\ee
The conserved momenta give
\bea
&& {r^3\over A} = {Qr^4 (\del_rx^+ + {r_0^4\over 2Q} \del_rx^-)-\del_rx^-
\over\sqrt{{1\over 1-r_0^4r^4}
- 2(\del_rx^+)(\del_rx^-) + Qr^4 (\del_rx^+ + {r_0^4\over 2Q} \del_rx^-)^2}}\ ,
\nonumber\\
&& {r^3\over AB} = {\del_rx^+-{r_0^4r^4\over 2}
(\del_rx^+ + {r_0^4\over 2Q} \del_rx^-)\over\sqrt{{1\over 1-r_0^4r^4}
- 2(\del_rx^+)(\del_rx^-) + Qr^4 (\del_rx^+ + {r_0^4\over 2Q} \del_rx^-)^2}}\ ,
\eea
This gives
\bea\label{delrx+-}
&& \del_rx^-={Qr^4-B+{Br_0^4r^4\over 2}\over 1-r_0^4r^4}
{1\over \sqrt{{A^2B^2\over r^6} (1-r_0^4r^4) +
Qr^4(1+{r_0^4\over 2Q}B)^2-2B}}\ ,\nonumber\\
&& \del_rx^+={1-{r_0^4r^4\over 2}-{Br_0^8r^4\over 4Q}\over 1-r_0^4r^4}
{1\over \sqrt{{A^2B^2\over r^6} (1-r_0^4r^4) +
Qr^4(1+{r_0^4\over 2Q}B)^2-2B}}\ .
\eea
Then we have
\bea\label{Deltax+-}
&& \Delta x^- = \int_0^{r_*} dr {Qr^4-B+{Br_0^4r^4\over 2}\over 1-r_0^4r^4}
{1\over \sqrt{{A^2B^2\over r^6} (1-r_0^4r^4) +
Qr^4(1+{r_0^4\over 2Q}B)^2-2B}}\ ,\nonumber\\
&& \Delta x^+ = \int_0^{r_*} dr
{1-{r_0^4r^4\over 2}-{Br_0^8r^4\over 4Q}\over 1-r_0^4r^4}
{1\over \sqrt{{A^2B^2\over r^6} (1-r_0^4r^4) +
Qr^4(1+{r_0^4\over 2Q}B)^2-2B}}\ .
\eea
The entanglement entropy finally becomes
\be\label{AEQr0}
S_A\ \sim\ V_2 \int_\epsilon^{r_*} {dr\over r^3}
{AB\over\sqrt{A^2B^2 - A^2B^2r_0^4 r^4 - 2Br^6
+ (1+{r_0^4\over 2Q}B)^2 Qr^{10}}}
\ee
For $r_0=0$ and $Q=0$, these expressions reduce to those for $AdS_5$.\
In the limit $r_0\ra 0$ with $Q$ fixed, these expressions can be
seen to reduce to (\ref{delx}) (\ref{SEadspw}) for the $AdS_5$ plane wave.

For the unboosted black brane $Q={r_0^4\over 2}$ with $x^\pm$
appearing symmetrically, we can set $B=1$: this then gives\
$\del_r x^-= {-r^3\over\sqrt{(A^2-2r^6)(1-r_0^4 r^4)}} = -\del_r x^+ ,
\ S_A\ \sim\ V_2 \int_0^{r_*} {dr\over r^3}
{A\over \sqrt{(A^2-2r^6)(1-r_0^4 r^4)}}$ , giving
$\del_r t={1\over\sqrt{2}} (\del_rx^++\del_rx^-)=0$ and the expected
bulk minimal surface on a constant time slice. The denominator in
these expressions is a factorized limit of those in (\ref{delrx+-}),
(\ref{Deltax+-}), (\ref{AEQr0}).

More generally, let us consider the boosted black brane with
$\lambda^2={2Q\over r_0^4}\neq 1$: the $AdS$ plane wave arises in
the extreme limit of infinite boost $\lambda\ra\infty$.
Then it can be seen that the denominator in
(\ref{delrx+-}), (\ref{Deltax+-}), (\ref{AEQr0}), can always
be factorized as in the unboosted case above if
$B={2Q\over r_0^4}=\lambda^2$.\
%\ (in the $AdS_5$ plane wave limit, we have $B\ra\infty$).\
At this factorization point, we have\ \
$\Delta x^- = \int_0^{r_*} dr {(-2Q/r_0^4)
\over \sqrt{({A^2B^2\over r^6}-2B) (1-r_0^4r^4)}}$\ ,\ \
$\Delta x^+ = \int_0^{r_*} dr
{1\over \sqrt{({A^2B^2\over r^6}-2B) (1-r_0^4r^4)}}$~,\ with \
$S_A\ \sim\ V_2\int {dr\over r^3}
{AB\over\sqrt{(A^2B^2 - 2Br^6) (1-r_0^4r^4)}}$~.\ This gives\
$\Delta x^-=-\lambda^2 \Delta x^+$,\ \ie\ \
$\sqrt{2} \Delta t=\lambda \Delta x^+ + \lambda^{-1}\Delta x^- = 0$~.
Furthermore, we have\
$\sqrt{2} \lambda \del_r t = \del_r x^- + \lambda^2\del_r x^+\ \propto\
(-B+{2Q\over r_0^4}) = 0$,\ \ie\ we have a constant-$t$ bulk minimal
surface. Here $t = {\lambda x^+ + \lambda^{-1} x^-\over\sqrt{2}}$ is
the unboosted time coordinate: recall that the
$[x^\pm\leftrightarrow (t,x_3)]$-coordinate-transformation maps the
boosted $AdS$ black brane (\ref{blackD3lightcone0}) to an unboosted
brane.
Thus the factorization point $B={2Q\over r_0^4}$ is simply the case
where $\del_rt=0$ for the extremal surface, \ie\ the surface is at
constant-$t$. This is of course the usual connected constant-time
extremal surface in the $AdS$ black brane corresponding to the
unboosted observer, rewritten in $x^\pm$-coordinates (with the
momentum $B$ tuned as above). From the point of view of the boosted
brane, this is not the most natural: $T= {x^+ + x^-\over\sqrt{2}}$
is the more natural time coordinate for the boosted case.

With a view to obtaining large values for $\Delta x^\pm$ \ie\ large
size, we note that the parameters $A,B$ can be tuned to a location
where there is a double zero of the denominator surface,
giving $\Delta x^\pm\ra\infty$: this happens when
\bea\label{doublezero}
&& V(r_*)\equiv A^2B^2 - A^2B^2r_0^4 r_*^4 - 2Br_*^6
+ (1+{r_0^4\over 2Q}B)^2 Qr_*^{10} = 0\ ,\nonumber\\
&& V'(r_*) = -2A^2B^2r_0^4r_*^3 - 6Br_*^5
+ 5 (1+{r_0^4\over 2Q}B)^2 Qr_*^9 = 0\ .
\eea
A rough plot of this surface $V(r)$ suggests that the nature of the
surface does not change much for small $r_0$. In particular, we can
expand these expressions in powers of $r_0^4$ to obtain the leading
corrections
\be\label{B-Adspwr0corr}
B = {5\over 6} Q r_*^4 + {17\over 36} Qr_0^4 r_*^8 + \ldots \ , \qquad\quad
A^2 = {24\over 25} {r_*^2\over Q} + {4\over 125} {r_0^4 r_*^6\over Q}
+ \ldots\ .
\ee
So it appears that the double zero location shifts a little from the
$AdS_5$ plane wave values (\ref{doublezeroAdSpw}) but continues to
exist with small $r_0$-dependent corrections: $\Delta x^\pm$ are
positively divergent.
%Thus we see that the value $B_-$ gives a double zero location
%at which $\Delta x^\pm$ diverge but as in the $AdS$ plane wave
%(\ref{doublezeroAdSpw}), with small $r_0$
This is perhaps not surprising since the scale $Q^{1/4}$ governing
the $AdS$ plane wave is widely separated from the horizon scale
$r_0$ in the near-extremal limit with small $r_0$.

%As we increase ${r_0^4\over 2Q}$ this behaviour might change to reflect
%the black D3-brane behaviour, which admits no such double zero location.
More generally, eliminating the parameter $A$ in (\ref{doublezero}),
we obtain
\be
%&& \qquad\qquad\qquad\qquad
A^2B^2 = {(2Br_*^6-(1+{r_0^4\over 2Q}B)^2 Qr_*^{10})\over 1-r_0^4r_*^4}\ ,
%\nonumber\\
%&& {r_*^2\over 4Q (1-r_0^4r_*^4)} \Big( B^2 r_0^8 r_*^4 (5-3r_0^4r_*^4)
%- 4BQ (6-7r_0^4r_*^4+3r_0^8r_*^8) + 4Q^2r_*^4 (5-3r_0^4r_*^4)\Big) = 0\ .
%\qquad
\ee
and a quadratic equation for $B$ which is solved as
\be\label{BQr0}
B_- = \left({2Q\over r_0^4}\right) {6-7r_0^4r_*^4+3r_0^8r_*^8\ -\
2\sqrt{3} (1-r_0^4r_*^4)\sqrt{3-r_0^4r_*^4}\over
r_0^4r_*^4(5-3r_0^4r_*^4)}\ .
\ee
Expanding about $r_0^4r_*^4\sim 0$, we see that this root matches
the $AdS$ plane wave values with the leading $r_0$-corrections
(\ref{B-Adspwr0corr}), while the other root $B_+$ diverges in the
$AdS$ plane wave limit ($r_0\ra 0$) and is discarded.
This admits a real solution and thus the double zero location exists
only if the discriminant is positive, \ie\
%3-7r_0^4r_*^4+5r_0^8r_*^8-r_0^{12}r_*^{12}  %\be
%(6-7r_0^4r_*^4+3r_0^8r_*^8)^2 -  r_0^8r_*^8 (5-3r_0^4r_*^4)^2
%= (1-r_0^4r_*^4)^2 (3-r_0^4r_*^4) \geq 0\ .
%\ee
\be
r_*^4 \leq {3\over r_0^4}\ .
\ee
Note that this location is inside the horizon at $r_0^4r^4=1$.
%Exploring this further, we see that (\ref{BQr0}) expands for
%$r_0^4r_*^4\sim 0$ (small width) as
%\be\label{BQr0-r0rt0}
%B_- = {2Q\over r_0^4} \left( {5r_0^4r_*^4\over 12} + {17 r_0^8r_*^8\over 72}
%\right) ,\quad {\rm or}\quad
%B_+ = {2Q\over r_0^4} \left( {12\over 5r_0^4r_*^4} - {34\over 25} -
%{49r_0^4r_*^4\over 1500} \right) .
%\ee
%The second solution while the first solution

One might have worried that the phase transition we observed is an
artifact due to the singularity at the IR limit $r\to \infty$.
%To answer this problem, here we would like to
Regulating the AdS plane wave geometry by placing an event horizon in
the IR, %following \cite{Narayan:2012hk}
we have calculated the holographic entanglement entropy (HEE) to see
if there is any problem. This regularized $AdS$ plane wave is obtained
by boosting the $AdS$ Schwarzschild black brane: the
$x^\pm$-coordinates appear in some ways inconvenient calculationally
from the point of view of identifying the connected bulk extremal
surface for large subsystem size as appropriate for the boosted
observer. Instead it turns out to be more convenient to use different
coordinates as we will describe below: we will later map this to
the present discussion.

\subsection{Regularized AdS Plane Wave: other coordinates}

Consider the HEE in the regularized AdS plane wave
for the strip subsystem (width $l$) in the case B. This
is equivalent to the HEE in the AdS Schwarzschild black hole \be ds^2=\f{dr^2}{r^2(1-r_0^4
r^4)}-\f{1-r^4r_0^4}{r^2}dt^2+\f{dy^2+\sum_{i=1}^{d-2}dx_i^2}{r^2},
\label{bhmet} \ee for the strip subsystem defined by the interval
$(\Delta X^+,\Delta X^-)$. They are related by \be
Q=\f{r_0^4\lambda^2}{2},\ \ \ \Delta X^{+}=\f{\lambda l}{\s{2}},\ \
\Delta X^{-}=-\f{l}{\s{2}\lambda},\ee where $X^\pm=\f{t\pm
y}{\s{2}}$. Below we focus on the $d=4$ case.

By solving the equations for the extremal surface $\gamma_A$ as before, in the end we find that
the size of the strip is given by \ba && \Delta
y=2\int^{r_*}_{\ep}dr\f{r^3}{\s{(a^2b^2-r^6)(1-r_0^4r^4)+b^2r^6}},\label{dy}\\
&& \Delta
t=2\int^{r_*}_{\ep}dr\f{-br^3}{(1-r_0^4r^4)\s{(a^2b^2-r^6)(1-r_0^4r^4)+b^2r^6}}
\label{dt} , \ea where $r_*$ is defined by \be
(a^2b^2-r_*^6)(1-r_0^4r_*^4)+b^2r_*^6=0. \label{condaf} \ee
The area of $\gamma_A$ is written as \be
\mbox{Area}=2\int^{r_*}_{\ep}\f{dr}{r^3}\cdot\f{ab}{\s{(a^2b^2-r^6)(1-r_0^4r^4)+b^2r^6}}.\label{areaexp}
\ee
In order to keep the inside of the square root i.e. the function
\be
g(r)=(a^2b^2-r^6)(1-r_0^4r^4)+b^2r^6,\label{fgn}
\ee
positive, we need to require
\be
|b|<\f{\s{3}(1-r_0^4r_*^4)}{\s{3-r_0^4r_*^4}}, \label{bcond}
\ee
which is equivalent to $g'(r_*)<0$.
For simplicity, we will set $r_0=1$ without losing the generality. If we define $r=r_*$ to be the turning point, we have $g(r_*)=0$ and therefore we find
\be
a^2b^2=\f{r_*^6(1-r_*^4)-r_*^6 b^2}{1-r_*^4}(\geq 0).
\ee

In this regularized geometry, there is no disconnected solution as it cannot
penetrate the horizon. This guarantees completely well-controlled calculations.
Then we need to ask if we can smoothly take the limit $\lambda\to \infty$, where the horizon is pushed into the deep IR and the AdS plan wave is recovered. To see this, it is crucial to check if we can realize the extremal surface $\gamma_A$ corresponds to any arbitrary large values of $\Delta y$ and $\Delta t$ with the space-like condition $|\Delta y|>|\Delta t|$).

Thus let us study the limits where $\Delta y$ and $\Delta t$ get divergent. They are given by
some combinations of (i) the limit $r_*\to 1$, where $\gamma_A$ approaches to the horizon, and (ii) the condition where $g(r)$ develops the double zero.
Thus first we define an infinitesimally small parameter by $\eta=1-r_*>0$.
  The double zero condition is equivalent to the condition $g'(r_*)=0$ and this fixes $b$ in terms of
$r_*$ as follows:
\be
b=\f{\s{3}(1-r_*^4)}{\s{3-r_*^4}}\simeq 2\s{6}\eta+O(\eta^2).
\ee
Therefore we can take the second infinitesimally small parameter $\epsilon$ to be
\be
\ep=-b+\f{\s{3}(1-r_*^4)}{\s{3-r_*^4}}\simeq -b+ 2\s{6}\eta.
\ee
So we have {\it two} infinitesimally small positive parameters $\eta>0$ and $\ep>0$ which control the limits.

Now we define the near horizon coordinate by
\be
\delta=r_*-r.
\ee
Then we can expand $g(r)$ near the turning point $r=r_*$ as follows
\ba
&& g(r)=c_1\ep\delta +c_2\delta^2+ O(\delta^3), \no
&& c_1=4\s{6}+O(\eta)+O(\ep),\no
&& c_2=24+O(\eta)+O(\ep).
\ea
It is also useful to note
\be
1-r^4=4\delta +4\eta +O(\eta^2).
\ee

We can estimate (\ref{dy}) and (\ref{dt}) by using the above approximation.
This leads to
\ba
&& \Delta y\sim 2\int^{1}_{0}d\delta \f{1}{\s{c_1\ep\delta+c_2\delta^2}}=-\f{4}{\s{c_2}}\cdot\log \left[\f{\s{c_1c_2\ep}}{c_2+\s{c_2(c_2+c_1\ep)}}\right],\no
&& \Delta t\sim  -2b\int^{1}_{0}d\delta \f{1}{4(\delta+\eta)\s{c_1\ep\delta+c_2\delta^2}}
=-\f{(2\s{6}\eta-\ep)}{\s{\eta(c_2\eta-c_1\ep)}}\cdot \mbox{arctanh} \left[\f{\s{c_2\eta-c_1\ep}}{\s{\eta(c_2+c_1\ep)}}\right].\no
\ea
Notice that since $b\geq 0$, we need to require
\be
2\s{6}\eta-\ep\geq 0.\label{limr}
\ee

We can think of many different limits, where both $\ep$ and $\eta$ go to zero. It is always true that $\Delta y$ gets divergent as
\be
\Delta y\simeq -\f{1}{\s{6}}\log \ep.
\ee

Now we take the following limit
\be\label{lim-scaling}
\ep\sim \eta^k\to 0,
\ee
with $k\geq 1$. This leads to the divergent $\Delta t$ as follows
\be
\Delta t\simeq \f{1}{2}\log\f{\ep}{\eta} \simeq\f{k-1}{2k}\log \ep,
\ee
where we have employed the expansion: arctanh$(1-x)\simeq -\f{1}{2}\log(x/2)+O(x)$ in the
limit $x\to 0$.
Thus we find
\be
\f{|\Delta t|}{|\Delta y|}=\f{\s{6}}{2}\cdot \f{k-1}{k}.
\ee
This shows that we can construct a connected extremal surface which leads to $\Delta y\to \infty$ (i.e. infinitely large width) for any ratio
which satisfies
\be
0<\f{|\Delta t|}{|\Delta y|}<\f{\s{6}}{2}.
\ee
This covers all regions we wanted. In this way, we can conclude that the AdS plane wave limit $\lambda \to \infty$ is smooth.

To compare with our previous discussions in the $x^\pm$-coordinates,
we can map the coefficients $A,B$ in (\ref{AEQr0})) and $a,b$ in
(\ref{areaexp}): this gives
\be
b = {1-{B\over\lambda^2}\over 1+{B\over\lambda^2}}\ ,
\ee
and correspondingly for the parameter $a$. The double zero location
$B_-$ in (\ref{BQr0}) then maps to
\be
b = {1-{B\over\lambda^2}\over 1+{B\over\lambda^2}}
= {\sqrt{3} (1-r_0^4r_t^4)\over \sqrt{3-r_0^4r_t^4}}\ ,
\ee
in agreement with (\ref{condaf}), (\ref{bcond}). This then recasts
(\ref{Deltax+-}) as
\bea
 \Delta x^- = \int {\lambda^2\ ({r_0^4r^4\over 2} (1+{B\over\lambda^2})-
{B\over\lambda^2})\over (1-r_0^4r^4)\sqrt{\ldots}} =
\int {\lambda^2\ (b - (1-r_0^4r^4)) \ dr\over (1+b)(1-r_0^4r_t^4)
\sqrt{{a^2b^2\over r^6} (1-r_0^4r^4) - (1-b^2) + r_0^4r^4}}\ ,
\nonumber\\
 \Delta x^+ = \int {1-{r_0^4r^4\over 2} (1+{B\over\lambda^2})\over
(1-r_0^4r^4)\sqrt{\ldots}} =
\int { (b + (1-r_0^4r^4))\ dr\over (1+b)(1-r_0^4r_t^4)
\sqrt{{a^2b^2\over r^6} (1-r_0^4r^4) - (1-b^2) + r_0^4r^4}}\ .\ \ \
\eea
We then have
\be
\Delta x^- = -{\lambda^2\over 1+b} (\Delta t + \Delta y)\ ,\qquad
\Delta x^+ = -{1\over 1+b} (\Delta t - \Delta y)\ .
\ee
% \Delta t = -{(1+b)\over 2} ({\Delta x^-\over\lambda^2}+\Delta x^+)\ ,\qquad
%\Delta y = {(1+b)\over 2} (\Delta x^+-{\Delta x^-\over\lambda^2})\ .
A constant $t$ surface has $\Delta x^-=-\lambda^2 \Delta x^+$, as expected.
Scaling to the double zero $b\ra 0$, and to the horizon $r_0r_*\ra 1$,
with $b$ scaling faster as in (\ref{lim-scaling}), we see that
$\Delta x^\pm$ acquire log-divergences\
%\be
$\Delta x^+\ra k_1 \log (..) ,\ \Delta x^-\ra -k_2 \lambda^2 \log (..)$ ,
%\ee
corresponding to a spacelike surface as long as the constants
$k_1,k_2$ are not identical.\ The $AdS$ plane wave calculation can be
defined as the limit $r_0\ra 0,\ r_*\ra\infty,\ r_0r_*\ra 1$ with the
scaling above: this is effectively equivalent to the disconnected
surface mentioned earlier. Thus the overall picture of the extremal
surfaces appears consistent.

\section{Conclusions}

$AdS$ plane waves arise as normalizable null deformations of
$AdS\times S$ spaces \cite{Narayan:2012hk,Narayan:2012wn}: they are
dual to excited CFT states with lightcone momentum density $T_{++}$
turned on. These spacetimes are likely $\al'$-exact string backgrounds
similar to plane waves and $AdS_5\times S^5$. With $T_{++}\sim Q$
uniform, these spacetimes are spatially homogenous and can be obtained
as certain limits of boosted $AdS$ Schwarzschild black branes
\cite{Singh:2012un}. Thus they are among the simplest anisotropic and excited
systems in holographic setups. The $AdS_5$ plane wave upon
$x^+$-dimensional reduction gives a background lying in the
hyperscaling violating family with ``$\theta=d-1$'', exhibiting
logarithmic behavior of entanglement entropy
holographically. In the higher dimensional description, the
corresponding extremal surface lies on a constant lightcone time $x^-$
slice and extends along the $x^+$-direction.

In this paper, we have explored holographic entanglement entropy (HEE)
for these $AdS_{d+1}$ plane waves for strip-shaped (spacelike)
subsystems.  There are two different cases: case A and case B as
explained in Fig.\ref{fig:sub}. In both case, the leading divergent
contribution to HEE is the familiar area law term. The finite part
behaves differently in the two cases.

When the strip is parallel to the energy flux (case A), the finite
part of HEE is a monotonically increasing function of the width $l$ of
the strip. Though it is smaller than the extensive thermal entropy, it
is always greater than the HEE for the pure AdS dual to the ground
state. In the $AdS_5$ plane wave case, this finite piece grows
logarithmically as\ $S^{finite}_{A}\sim V_2 \sqrt{Q} \log(l Q^{1/4})$
with size $l$ and is perhaps a finite reflection of the log-growth in
the null case above. It was argued in \cite{OTU,Sa} that hyperscaling
violating metrics with ``$\theta=d-1$'' are dual to systems with Fermi
surfaces. Indeed, if we regard $Q^{1/4}$ as the scale of fermi
momentum (~energy) $k_F$, then $S^{finite}_{A}$ agrees with the
expected behavior in systems with fermi surfaces \cite{OTU}. It is
therefore interesting to obtain a deeper understanding of the origin
of the logarithmic behavior in the $AdS_5$ plane wave backgrounds from
a field theory point of view. In particular, one would like to
understand if this behaviour is due to a Fermi surface or some
alternative mechanism.
In the $AdS_4$ plane wave case, the finite part grows with $l$ like
$S_A^{finite}\sim V_1 \sqrt{l} \sqrt{Q}$, so that the logarithmic term in the $AdS_5$ case is replaced by 
the power,  $\sqrt{l}$.
To the best of our knowledge, the $AdS_4$ plane wave is the first known  example of a construction 
arising from  string theory which gives rise to such a power law enhancement. 

On the other hand, when the
strip is orthogonal to the energy flux (case B), we find a phase
transition such that for $l$ larger than $l_c\sim Q^{-1/d}$, the HEE
becomes constant. Using the regulated description in
terms of boosted $AdS$ Schwarzschild black branes with horizon at
$r_0$, we have seen that this behaviour persists for small $r_0$:
this is expected since the scale $Q$ is more dominant here. For
general boost, the connected extremal surface can be identified in
a certain scaling limit where the surface approaches the horizon
(and develops a double zero).

One possible intuitive explanation of these different behaviors in case A and case B is as follows. Since the system we consider
has the energy flux $T_{++}\sim Q$, each excited mode has the wave length of order $(T_{++})^{1/d}\sim Q^{1/d}$ along the energy flux direction. Therefore in case B, where
the HEE measures the correlation in the flux direction, the HEE gets trivial
(i.e. constant) for
a large width $l>l_c\sim Q^{1/d}$ because we can regard $l_c$ as the correlation length.
This is the reason why we found the phase transition phenomenon in our holographic analysis of case B. On the contrary, in the direction orthogonal to the flux, the situation is similar to
the ground state and the correlation length should be infinite. Therefore we do not have
any phase transition in case A. It is very interesting to analyze the same question from CFT calculations in e.g. free field theories.

\vspace{5mm}

\noindent {\bf Acknowledgements:}\ TT thanks I. Klebanov, A. Lawrence,
S. Ryu for useful discussions.  We thank the Organizers of the String
Theory Discussion Meeting, June '12, at the International Center for
Theoretical Sciences (ICTS), Bangalore, where this work began. KN and
SPT thank the Organizers of the Indian Strings Meeting (ISM2012),
Puri, India for hospitality at the final stages of this work. TT is
grateful to the organizers and to the Princeton Center for Theoretical
Science for the workshop ``Entanglement in Discrete and Continuous
Quantum Systems,'' where the result of this work was presented.  SPT
would like to acknowledge the hospitality of the Yukawa Institute Of
Theoretical Physics during his visit to attend the workshop on
"Gauge/Gravity Duality'' where some of this research was done.  KN is
partially supported by a Ramanujan Fellowship, Department of Science
and Technology, Govt of India.  TT is supported by JSPS Grant-in-Aid
for Challenging Exploratory Research No.24654057. TT is partially
supported by World Premier International Research Center Initiative
(WPI Initiative) from the Japan Ministry of Education, Culture,
Sports, Science and Technology (MEXT). SPT acknowledges support from
the J. C. Bose Fellowship, Department of Science and Technology, Govt
of India.

%\newpage

%\appendix
%\section{Useful Formula} In this subsection, we summarize some

\end{document}